\PassOptionsToPackage{cmyk,table}{xcolor}
\documentclass[conference,pbalance]{iaria} 
\pdfoutput=1 

\usepackage{silence}
\usepackage{amsmath}
\usepackage{amssymb}
\usepackage{graphicx}
\usepackage{algorithm}
\usepackage{algpseudocode}
\usepackage{hyperref}

\usepackage{placeins}
\usepackage{makecell}
\usepackage{booktabs}

\usepackage[babel=true,english=american]{csquotes}
\usepackage[english]{babel} 

\usepackage[shortcuts]{extdash} 

\usepackage[alwaysadjust]{paralist}

\usepackage[font=footnotesize]{subfig}

\usepackage{fontawesome} 
\usepackage{relsize}     

\usepackage{xcolor}
\usepackage{soul}

\definecolor{revisionyellow}{RGB}{255,245,170}
\definecolor{revisionblue}{RGB}{0,70,160}

\sethlcolor{revisionyellow}

\newif\ifshowchanges
\showchangestrue  

\usepackage{xcolor}
\usepackage{soul}

\definecolor{revisionyellow}{RGB}{255,245,170}
\sethlcolor{revisionyellow}

\soulregister\cite7
\soulregister\textcite7
\soulregister\parencite7
\soulregister\ref7
\soulregister\autoref7

\newif\ifshowchanges
\showchangesfalse  

\ifshowchanges
    \newcommand{\rev}[1]{\hl{#1}}
    \newcommand{\revcomplex}[1]{\hl{#1}}
    \newcommand{\revheading}[1]{%
        \texorpdfstring{\colorbox{revisionyellow}{#1}}{#1}%
    }
\else
    \newcommand{\rev}[1]{#1}
    \newcommand{\revcomplex}[1]{#1}
    \newcommand{\revheading}[1]{#1}
\fi

\title{A Framework to Quantify the Probability of Future Cyber Loss Events}

\author{
    \IEEEauthorblockN{
        Siem Peters\,\orcidlink{0009-0003-3706-9626}
    }
    \IEEEauthorblockA{
        Mnemonic\\
        Oslo, Norway\\
        e-mail: \texttt{siemp@mnemonic.no}
    }

    \and

    \IEEEauthorblockN{
        Martin Eian\,\orcidlink{0009-0004-7461-3202}
    }
    \IEEEauthorblockA{
        Mnemonic\\
        Oslo, Norway\\
        e-mail: \texttt{meian@mnemonic.no}
    }
}

\begin{document}

\maketitle

\begin{abstract}

Cybersecurity risk quantification remains challenging due to limited operational data and difficulties in quantifying Loss Event Frequency (LEF). This paper introduces the Loss Event Frequency Security Analyser (LEFSA), a probabilistic framework that reformulates LEF estimation as machine-level \rev{Cyber Loss Event} (CLE) prediction combined with hierarchical infrastructure-level aggregation. LEFSA estimates calibrated machine-level CLE probabilities from operational cybersecurity telemetry and aggregates them across infrastructure layers while accounting for machine-level dependencies. This provides a foundation for scalable, explainable, and operationally applicable cyber risk estimation at the level of machines, services, business processes, and the entire organization. The framework was evaluated using proprietary Managed Detection \& Response telemetry from 23 organizations using Microsoft Defender for Endpoint. \rev{XGBoost achieved the strongest predictive performance, with a mean area under the receiver operating characteristic curve of 0.90} and consistently low calibration error across evaluation periods. The results demonstrate that operational cybersecurity telemetry contains substantial predictive information for future CLE occurrence, supporting probabilistic machine-level modeling and hierarchical aggregation as a promising foundation for quantitative, data-driven cyber risk management.

\end{abstract}


\section{Introduction}\label{sec:intro}

The number of cyber incidents has increased significantly in recent years, with malicious attacks nearly doubling compared to pre-pandemic levels \cite{GlobalFinancialStabilityReportApril2024}. Industry data further indicate an increasing frequency and severity of \rev{Cyber Loss Events} (CLEs), particularly for large losses \cite{chubb-2025,allianz-2024}, while reported cyber insurance claims increased by roughly 40\% year-over-year in 2024 \cite{naic-2025}. Recent advances in AI are accelerating the cyber attack lifecycle, from vulnerability discovery to exploitation, increasing the need for proactive and data-driven cybersecurity risk management frameworks \cite{kela-cyber-threat-intelligence-2025}.

Cybersecurity risk refers to the potential for financial loss, operational disruption, or reputational damage arising from failures in an organization’s \rev{Information Technology} (IT) systems caused by cyberattacks \cite{florackis}. While modeling the monetary impact of CLEs has been widely studied, particularly from an insurance perspective \cite{He_Jin_Li_2024,Carannante_Mazzoccoli_2025}, estimating the likelihood of a CLE affecting IT infrastructure remains largely qualitative in both academia and industry.

\subsection{\revheading{State-of-the-art}}

\revcomplex{Established information security risk-assessment frameworks and methodologies include ISO/IEC 27005, NIST SP 800-30, OCTAVE, and CORAS, which provide structured processes for identifying, analysing, and treating cyber risks \cite{sanchez-garcia-2022,ISO-2022,nist-2012,octave-1999,lund-2010}. While these approaches offer well-established guidance for risk assessment, they do not prescribe a single probabilistic or data-driven method for estimating loss frequencies or financial loss magnitudes from operational data.}

\revcomplex{This study therefore builds on Factor Analysis of Information Risk (FAIR) \cite{freund-2025}, one of the most established quantitative cyber-risk frameworks.} \rev{FAIR is particularly suitable because it expresses cyber risk quantitatively through two components}, \rev{namely the frequency and monetary impact of loss events associated with defined cyber-risk scenarios}, combined representing the cyber risk of an organization. The advantages of FAIR are (1) a structured and consistent methodology, (2) quantitative (financially grounded) output, (3) transparency, and (4) compatibility with probabilistic methods.

The Loss Event Frequency (LEF) component of FAIR is the estimated rate at which a specific threat scenario is expected to result in a loss event over a given time period. It represents the likelihood dimension of risk and is calculated as the combination of threat event frequency and susceptibility. \revcomplex{In operational settings, modeling the LEF component often proves challenging due to (1) limited relevant and high-quality scenario-specific data and (2) difficulties in operationalizing and estimating contact frequency and susceptibility. Moreover, while FAIR defines the relevant factors and their conceptual relationships, it does not prescribe a complete empirical procedure for acquiring organization-specific data, selecting and validating probability distributions, or learning these relationships from operational telemetry.}

\revcomplex{Related work has extended FAIR using Bayesian networks to support more flexible distributions, dependencies, and model structures \cite{wang-2020}. Although this relaxes some modelling restrictions within the FAIR framework, it still relies on scenario-level inputs that may be unavailable or difficult to estimate and validate from operational data.}

Many existing quantitative cyber-risk prediction approaches operate at the organization level using externally observable characteristics and public data \cite{Liu_Sarabi_Zhang_Naghizadeh_Karir_Bailey_Liu_2015,Okutan_Werner_Yang_McConky_2018}. \revcomplex{\textcite{leslie-2017} instead used aggregated internal Managed Security Service Provider (MSSP) data to predict organization-level intrusion counts. Although these studies demonstrate that cyber incidents are predictable, organization-level models provide limited insight into which individual machines or infrastructure components generate the estimated risk.}

A more operationally applicable approach is to estimate future CLE probabilities directly at the machine level. Prior work shows that endpoint behaviour, software configuration, and file-appearance patterns can accurately predict future malware infections using internal telemetry \cite{Bilge_Han_DellAmico_2017,trendmicro,attack_prediction_survery_2025}. \rev{The reported predictive performance of these machine-level approaches is generally stronger than that of organization-level models, although differences in data and prediction tasks prevent direct comparison. However, they focus primarily on malware infection and do not provide a framework for predicting general CLEs, identifying their machine-level sources, and aggregating risk across all IT infrastructure layers.}

\subsection{\revheading{Contributions}}

This paper proposes \rev{Loss Event Frequency Security Analyser (LEFSA)}, a data-driven framework to estimate the likelihood of CLEs through machine-level CLE prediction and hierarchical infrastructure-level aggregation. Rather than modeling isolated machine-level outcomes or manually constructed threat scenarios, LEFSA estimates calibrated machine-level CLE probabilities and aggregates them across infrastructure layers. This enables probabilistic cyber risk estimation at the level of machines, services, business processes, and entire organization.

The main contributions of this paper are as follows:
\begin{itemize}

\item A reformulation of LEF estimation as a measurable machine-level probabilistic prediction problem, extending existing machine-level malware prediction approaches to generalized CLE estimation.

\item A hierarchical cyber risk aggregation framework that propagates machine-level CLE probabilities across infrastructure layers. This would enable cyber risk estimation at all IT infrastructure levels, like machines, services, business processes, or the entire organization.

\item A proposed probabilistic aggregation methodology for modeling predictive distributions of aggregated CLEs. This enables estimation of not only future expected CLEs, but also tail-risk outcomes and worst-case infrastructure-level cyber risks.

\item A discussion of both independent and correlated aggregation settings, including dynamic dependence structures and potential copula-based approaches for modeling joint CLE distributions.

\item An empirical evaluation using real-world cybersecurity data from multiple organizations.

\end{itemize}

\subsection{\revheading{Limitations}}

\rev{The scope of this work is limited to the estimation of LEF and does not include modeling the loss magnitude of cyber risk. The empirical evaluation uses proprietary Managed Detection \& Response (MDR) data that cannot be publicly shared; the dataset and experimental setting are described in Section \ref{Sec: Experiments}. Its empirical evaluation is limited to organizations covered by this dataset, and generalization to other environments depends on the availability and comparability of machine-level telemetry and alerts generated by intrusion detection platforms. The current aggregation procedure relies on simplifying assumptions regarding dependencies between machines.}

\rev{The remainder of this paper is organized as follows: Section \ref{Sec: LEFSA: A generalized loss event framework} describes the LEFSA framework for predicting and aggregating CLEs. Section \ref{Sec: Experiments} presents the experiments and results. Section \ref{Sec: Discussion} discusses the results, implications, and limitations, while Section \ref{Sec: Conclusion} concludes the paper and outlines future work.}

\section{LEFSA: A Generalized Loss Event Framework}
\label{Sec: LEFSA: A generalized loss event framework}

\subsection{Definitions}

There are $T$ time periods, indexed by $t \in \{ 1,\dots,T \}$. Let $\mathcal{M}$ denote the set of all machines, where each machine has a unique identifier. At each time $t$, only a subset $\mathcal{M}_t \subseteq \mathcal{M}$ is observed. The set of observed machines $\mathcal{M}_t$ may vary over time \rev{as machines are added, removed, or temporarily unobserved. $\mathcal{M}$ is expanded whenever a previously unseen machine appears.} For each machine $m \in \mathcal{M}_t$, let $X_{t,m} \in \{0,1\}$ denote the random variable indicating whether machine $m$ is involved in a CLE at time $t$. A CLE should be defined using measurable outcomes available in the data, such as malware detections or security incidents. $\mathcal{F}_t$ denotes the information set, \rev{i.e.,} the features available up to and including period $t$.

Organizations monitor a set of machines over time. This can be represented by the following matrix:
\[
\mathbf{X}_{1:T} =
\begin{bmatrix}
X_{1,1} & X_{1,2} & \dotsi & X_{1,M} \\
X_{2,1} & X_{2,2} & \dotsi & X_{2,M} \\
\vdots & \vdots & \ddots & \vdots \\
X_{T,1} & X_{T,2} & \dotsi & X_{T,M}
\end{bmatrix}
= \begin{bmatrix}
1 & - & 0 & \dotsi \\
0 & 1 & 0 & \dotsi \\
\vdots & \vdots & \vdots & \ddots
\end{bmatrix}
\]
where rows correspond to time periods $t=1,\dots,T$ and columns correspond to machines $m=1,\dots,M$. Missing entries indicate machines that were not observed during a given time period, reflecting that the observed machine set $\mathcal{M}_t$ may vary over time.

For infrastructure level $g$, define the aggregated random variable $Y_t^{(g)} = \sum_{m \in \mathcal{M}_{t,g}} X_{t,m}$, where $\mathcal{M}_{t,g}$ denotes the set of machines at level $g$. Then $Y_t^{(g)}$ represents the number of machines at infrastructure level $g$ affected by at least one CLE during period $t$.

\subsection{Machine-level LEF}

For every machine $m \in \mathcal{M}_t$, the conditional probability that machine $m$ will be involved in a CLE during the next time period $t+1$ is denoted as $p_{t+1,m} := \mathbb{P}(X_{t+1,m}=1|\mathcal{F}_t)$. To predict the probability of a future CLE on a single machine, a model is trained to estimate $p_{t+1,m}$.

The conditional probability is approximated by a model $f_{\hat{\theta}}$ that maps the information set $\mathcal F_t$ to the unit interval, \rev{i.e.,} $f_{\hat{\theta}}:\mathcal F_t \rightarrow [0,1]$, yielding the estimate $\hat p_{t+1,m}:=f_{\hat\theta}(\mathcal F_t)$.

To aggregate machine-level predictions, LEFSA requires calibrated probability estimates satisfying:
\begin{equation} 
    \label{Eq: True probability}
    \mathbb{P}(X_{t+1,m} = 1 | \hat{p}_{t+1,m}=p)=p \ \forall p \ \in [0,1],
\end{equation}
Such estimates may be obtained either directly from probabilistic models or, when necessary, through post-hoc calibration of score-based models.

\subsubsection{Probabilistic models}

Inherently probabilistic models, such as logistic regression, directly estimate the mapping $f_{\hat\theta}$ through likelihood-based inference. Under suitable conditions, these models may provide well-calibrated probability estimates.

\subsubsection{Score-based models}

More flexible machine learning models can offer improved predictive performance in complex or high-dimensional settings, although calibration may still need to be assessed empirically \cite{mizil-2005}. If the chosen model outputs a risk score rather than a calibrated probability satisfying (\ref{Eq: True probability}), the probability mapping $f_{\hat{\theta}}$ is implemented as a composition of two components:

\begin{enumerate}
    \item Let $h_{\hat{\theta}}$ denote the score-producing model, outputting a risk score:
    \[
        \hat{s}_{t+1,m} := h_{\hat{\theta}}(\mathcal{F}_t)
    \]

    \item A calibration map $v$ is then applied, where $v: \hat{s} \longmapsto [0,1]$, yielding
    \[
        \hat{p}_{t+1,m} := v(\hat{s}_{t+1,m}).
    \]
    Various calibration methods for $v$ are possible, such as Platt scaling \cite{platt-2000} or Isotonic regression \cite{zadrozny-2001}.
    
\end{enumerate}

\noindent
Machine-level LEF estimation involves: (1) defining a CLE, (2) constructing the information set $\mathcal{F}_t$, and (3) training and, if necessary, calibrating a model to estimate $\hat{p}_{t+1,m} \ \forall m \in \mathcal{M}_{t}$.

\subsection{Infrastructure-level LEF}

The second layer of LEFSA aggregates machine-level CLE probabilities across higher infrastructure levels, yielding aggregate predictive distributions over the number of CLEs within groups of machines.

Choose an infrastructure level of interest, \rev{e.g.,} a service or business process. Select the set of machines $m$ that support the chosen infrastructure level $g$, defined as $m \in \mathcal{M}_{t,g}$. The interest lies in modeling the future risk of a group of assets, \rev{i.e.,} $Y_{t+1}^{(g)} = \sum_{m \in \mathcal{M}_{t,g}} X_{t+1,m}$.

\subsubsection{Independent machines}

Although unlikely, the machine-level CLE variables may be conditionally independent given $\mathcal F_t$, \rev{i.e.,} $\operatorname{Cov}(X_{t+1,m},X_{t+1,m'} \mid \mathcal{F}_{t}) = 0$ for$\quad m \neq m'$. In that case, $Y_{t+1}^{(g)}$ follows a Poisson-binomial distribution with parameters $\{\hat{p}_{t+1,m}\}_{m \in \mathcal{M}_{t,g}}$. Note that the subscript of $\mathcal{M}$ is $t$ not $t+1$, since machine-level predictions are made for every running machine in the current time period $t$.

The Lindeberg–Feller Central Limit Theorem (CLT) for independent but non-identically distributed Bernoulli variables implies that
\[
\frac{Y_{t+1}^{(g)} - \hat{\mu}_{t+1}}{\hat{\sigma}_{t+1}} \;\Rightarrow\; \mathcal{N}(0,1),
\]
provided no single machine dominates the total variance \cite{billingsley-2012}. Therefore, a convenient approximation is to assume normality with mean $\hat{\mu}_{t+1} = \sum_{m \in \mathcal{M}_{t,g}} \hat{p}_{t+1,m}$ and variance $\hat{\sigma}_{t+1}^2 = \sum_{m \in \mathcal{M}_{t,g}} \hat{p}_{t+1,m}(1-\hat{p}_{t+1,m})$. 

Note that since CLEs are rare (see Section \ref{Sec: Experiments}), \rev{i.e.,} $\hat{p}_{t+1,m} \ll 1$, the Poisson-binomial distribution of $Y_{t+1}^{(g)}$ can alternatively be approximated by a Poisson distribution with rate $\lambda_{t+1} = \sum_{m \in \mathcal{M}_{t,g}} \hat{p}_{t+1,m}$.

\subsubsection{Correlated machines}

Machine-level CLEs may exhibit conditional dependence \cite{chen-2015}, \rev{i.e.,} $\operatorname{Cov}(X_{t+1,m},X_{t+1,m'} \mid \mathcal{F}_{t}) \neq 0$ for $m \neq m'$. In this case, it would require modeling the joint dependence structure between machines.

More generally, what must be modeled is the joint conditional distribution of the machine-level CLE vector $\mathbf{X}_{t+1}=\left( X_{t+1,m} : m \in \mathcal{M}_{t,g} \right)$, given the available information up to time $t$:

\begin{equation}
    F_{t+1}(\mathbf{x})=\mathbb{P}\left(\mathbf{X}_{t+1} \leq \mathbf{x}\mid\mathcal{F}_{t}\right)
\end{equation}
It contains both the marginal risk of each machine and the dependence structure between machines. The corresponding aggregate distribution is
\begin{equation}
    \mathbb{P}(Y_{t+1}^{(g)} = y \mid \mathcal{F}_{t})
    =
    \sum_{\mathbf{x}: \sum_{m \in \mathcal{M}_{t,g}} x_m = y}
    \mathbb{P}
    \left(
    \mathbf{X}_{t+1} = \mathbf{x}
    \mid
    \mathcal{F}_{t}
    \right)
\end{equation}
This shows that the distribution of $Y_{t+1}^{(g)}$ is fully determined by the joint distribution $F_{t+1}$. In this case, accurate aggregation requires not only predicting the machine-level future marginal CLE probabilities $\hat{p}_{t+1,m}$, but also the dependence structure between machines. Additionally, the data might exhibit time-varying dependence, \rev{i.e.,} $\mathbf{X}_{t+1} \mid \mathcal{F}_{t} \sim F_{t+1}$ where $F_{t+1}$ denotes a time-varying joint distribution.

Modeling dependence between machine-level CLEs is non-trivial, particularly for discrete and time-varying outcomes. One possible approach is to use copulas, which couple marginal distributions into a joint multivariate distribution \cite{thorsten-copulas}.

Copula-based approaches, including static vine copula-GARCH formulations and vine copula models with temporal and cross-group dependence structures, have been applied to cybersecurity risk data \cite{Peng_Xu_Xu_Hu_2018,li-2025}:
\begin{equation}
    F_{t+1}(x_1,\ldots,x_d)
    =
    C_{t+1}
    \left(
    F_{t+1,1}(x_1), \ldots, F_{t+1,d}(x_d)
    \right),
\end{equation}
where $C_{t+1}$ captures machine-level dependence and $d = |\mathcal M_{t,g}|$.

Time-varying dependence may be incorporated through dynamic copula parameters \cite{patton-2001}:
\begin{equation}
    F_{t+1}(x_1,\ldots,x_d)
    =
    C_{t+1}
    \left(
    F_{t+1,1}(x_1), \ldots, F_{t+1,d}(x_d);
    \theta_{t+1}
    \right)
\end{equation}
with
\begin{equation}
    \theta_{t+1}=h(\theta_t,Z_t,\varepsilon_{t+1}),
\end{equation}
where $Z_t$ may include cyber or infrastructure information such as vulnerability disclosures, patch status, threat activity, or network changes.

For short observation windows, CLEs may be modeled as binary variables $X_{t,m}\in\{0,1\}$, while longer windows may instead use count outcomes $X_{t,m}\in\mathbb N_0$. Although Sklar’s theorem guarantees a copula representation, uniqueness generally fails for discrete marginals, complicating both dependence interpretation and statistical inference \cite{genest-2007}.

In practice, dependence modeling with discrete marginals is often handled through approximation-based approaches, including latent continuous-variable constructions, Inference Functions for Margins (IFM), and simplified pair-copula or low-rank structures \cite{nikoloulopoulos-2013,joe-2014,panagiotelis-2016}. Extending these methods to dynamic settings remains an active research area and is discussed further in Section \ref{Sec: Discussion}.

Given marginal probabilities $\{\hat p_{t+1,m}\}_{m \in \mathcal{M}_{t,g}}$ and an estimated joint distribution $\widehat F_{t+1}$, the aggregate distribution of $Y_{t+1}^{(g)}$ can be approximated with Monte Carlo simulation. For $b = 1,\dots,B$, generate a correlated machine-level realization $\tilde{\mathbf X}_{t+1}^{(b)} \sim \widehat F_{t+1}$, where \rev{$\tilde{\mathbf X}_{t+1}^{(b)}=\left(\tilde X_{t+1,m}^{(b)}:m \in \mathcal M_{t,g}\right)$}. The corresponding aggregate realization is then

\[
\tilde Y_{t+1}^{(g,b)}
=
\sum_{m \in \mathcal M_{t,g}}
\tilde X_{t+1,m}^{(b)}.
\]
Repeating this procedure yields an empirical approximation of the predictive distribution of $Y_{t+1}^{(g)}$. Any higher-level aggregation, such as services, business processes, or organization-level risk, is performed directly from the estimated joint machine-level distribution $\widehat F_{t+1}$, thereby preserving the dependence structure across all machines and infrastructure layers.

\subsubsection{Risk metrics}

\begin{figure}[t]
    \centering
    \includegraphics[width=\columnwidth]{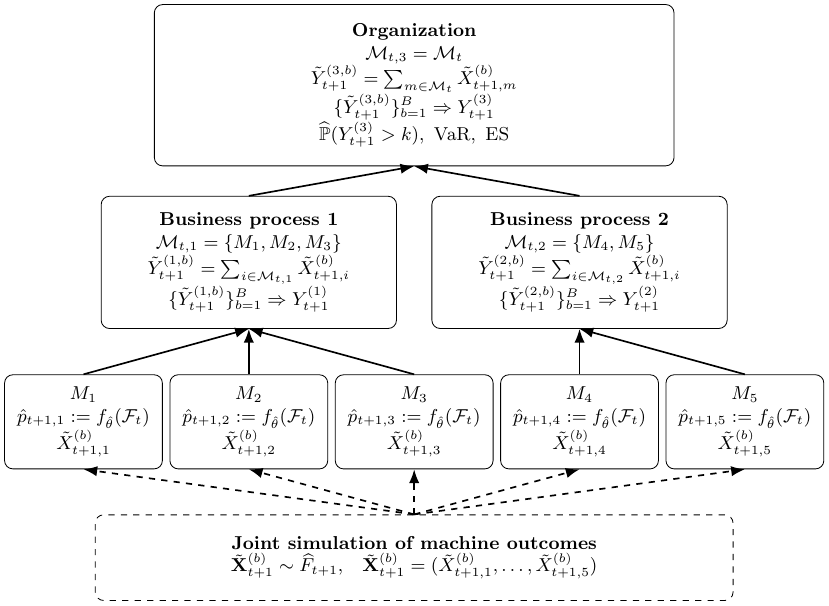}
    \caption{LEFSA aggregation framework. Machine-level CLE probabilities
    are combined through an estimated joint distribution and aggregated
    across infrastructure levels to obtain predictive risk distributions.}
    \label{fig:risk-aggregation}
\end{figure}

Having estimated the predictive distribution of $Y_{t+1}^{(g)}$, infrastructure-level cyber risk metrics can be computed. For example, if a business process requires four operational machines and maintains two backup servers, the probability of service disruption during period $t+1$ is $\mathbb P(Y_{t+1}^{(g)} > 2 \mid \mathcal F_t)$. More generally, the distribution of $Y_{t+1}^{(g)}$ enables estimation of Value-at-Risk (VaR), Expected Shortfall (ES), and other tail-risk measures (expressed in the number of CLEs). See Figure \ref{fig:risk-aggregation} for a complete overview of LEFSA.

\section{Experiments \& Results}\label{Sec: Experiments}

The LEFSA framework was evaluated using data from Mnemonic’s MDR service. Mnemonic AS is a Norwegian cyber security company, providing MDR and cyber risk services. Due to confidentiality agreements, the dataset cannot be released publicly. The empirical evaluation focuses primarily on machine-level LEF, while modeling time-varying machine dependencies for infrastructure-level LEF is left for future work.

\subsubsection{Data}

The dataset covers 110,924 unique machines from 23 organizations, all using \rev{Microsoft Defender for Endpoint (MDE)} \cite{microsoft_mde}. The data span 12 weeks, although not every machine is observed in every time period. The dataset consists of MDE telemetry, alerts generated by Mnemonic’s detection systems, and incident analysis results produced by Mnemonic’s Security Operations Center (SOC). The target variable is binary and indicates whether a machine was reported in a security incident. In total, the dataset contains 952,770 machine-period instances, representing unique combinations of machines and time periods. Of these, 0.134\% of the machines were reported in a security incident.

\subsubsection{Features}

We developed a pipeline to deploy LEFSA as a live service across 23 organizations. A preprocessing script transformed each observed machine into hourly machine profiles, represented as vectors of counts that capture machine behaviour and configuration during each hour. Following each time period, the hourly machine profiles were aggregated into a single feature vector per machine.

The selected features can be grouped into four categories, ordered from highest to lowest data availability: exposure, vulnerabilities, security controls, and threats. These are denoted by $\mathcal{F}_{t}=(\textbf{E}, \textbf{V}, \textbf{SC}, \textbf{T})$. In addition, the one-period lagged outcome \rev{$X_{t,m}$} is included as a predictor to capture temporal dependence. For machines inactive or not yet observed during period \rev{$t$}, the lagged value is set to zero.

Each training and testing instance consists of machine feature vectors during week $t$ and the corresponding target variable observed during the subsequent week $t+1$. Consequently, 12 weeks of observations yield 11 consecutive rolling-windows. The objective is to predict the probability of a CLE for machine $m$ during the subsequent time period conditional on the observed feature set, \rev{i.e.,} $\mathbb{P}(X_{t+1,m} = 1 \mid \mathcal{F}_{t})$.

\subsubsection{Model training \& tuning}

The first two rolling-windows were used for model selection and hyperparameter tuning. Logistic Regression, Random Forest, Histogram Gradient Boosting, and \rev{Extreme Gradient Boosting (XGBoost)} were trained and evaluated to identify the best-performing model. Among these, XGBoost achieved the strongest performance. XGBoost is a scalable and regularized gradient boosting framework based on decision trees that is designed for efficient and high-performance supervised learning tasks \cite{XGBoost}. Hyperparameter optimization was performed using Optuna \cite{optuna} to identify the optimal parameter configuration for the dataset.

The remaining rolling-windows were used for model evaluation. A rolling-window validation strategy was used, in which the model was iteratively trained on all available historical weeks and evaluated on the immediately following week. Specifically, for each iteration, the model was trained on weeks $1,\dots,t$ and evaluated on week $t+1$.

\subsubsection{Performance}

Table \ref{tab:performance_metrics} summarizes weekly predictive performance across the evaluation periods for models trained jointly on all organizations. Joint training generally outperformed organization-specific models. Alternative variants incorporating class weight balancing and calibration were evaluated, but did not further improve performance or calibration quality.

The model achieved strong discriminative performance, with \rev{a mean area under the receiver operating characteristic curve (ROC-AUC)} of 0.90 and accuracy above 99.8\% across all evaluation periods. Given the severe class imbalance (0.06\%-0.27\% positive observations per week), \rev{the area under the precision-recall curve (PR-AUC)} is also reported, yielding a mean value of 0.23. Relative to the weekly positive-class baseline, this corresponds to approximately 45-500 times the performance of a random classifier.

\begin{table}[htbp]
\centering
\caption{Summary of model performance metrics.}
\label{tab:performance_metrics}

\setlength{\tabcolsep}{3.5pt} 
\renewcommand{\arraystretch}{1.05}

\begin{tabular}{@{}lcccc@{}}
\toprule
\textbf{Metric} & \textbf{Mean} & \textbf{Median} &
\textbf{Min} & \textbf{Max} \\
\midrule
ROC-AUC           & 0.90    & 0.91    & 0.82    & 0.95    \\
PR-AUC            & 0.23    & 0.22    & 0.07    & 0.37    \\
Accuracy (\%)     & 99.87   & 99.91   & 99.75   & 99.95   \\
ECE (\%)          & 0.094   & 0.096   & 0.023   & 0.187   \\
Calibration slope & 1.00    & 1.00    & 0.72    & 1.30    \\
\bottomrule
\end{tabular}
\end{table}

\subsubsection{Calibration}

Table \ref{tab:performance_metrics} shows that the trained XGBoost model exhibited good probabilistic calibration, with consistently low \rev{Expected Calibration Errors (ECE)} and calibration slopes generally close to 1.0.

\subsubsection{Feature importance}

Feature importance was assessed using SHapley Additive exPlanations (SHAP) computed for each evaluation period. While the relative importance of individual features varied across weeks, several consistent patterns emerged. In particular, the lagged \rev{outcome variable $X_{t,m}$} was the most influential predictor for all evaluation periods.

\subsubsection{Machine dependencies}

Although machine-level predictions demonstrated good calibration, aggregate CLE counts remained substantially overdispersed relative to the independent Bernoulli assumption. Global calibration aligned the expected and observed CLE frequencies. However, substantial overdispersion persisted ($37.23$, $p<0.001$), indicating that the excess variability could not be explained by calibration error alone.

Within the LEFSA framework, this suggests that machine-level CLE variables might not be conditionally independent given the observed information set $\mathcal{F}_t$, \rev{i.e.,} $X_{t,m} \not\!\perp X_{t,m'} \mid \mathcal{F}_t$. Although unconditional pairwise correlations between machine-level CLE variables were generally small, substantial aggregate overdispersion remained after conditioning on the predicted probabilities. This indicates residual clustering, dependence structures, or shared latent risk factors between machines that are not fully captured by the machine-level prediction model.

\subsubsection{Computational requirements}

LEFSA was deployed on a server equipped with a 48-core Intel Xeon Gold 5118 CPU at 2.30\,GHz and 251\,GB of RAM. Table~\ref{tab:computational_costs} reports approximate operational resource requirements, normalized per 1,000 machines. LEFSA consists of two main computational components: (1) continuous stream processing of raw endpoint logs and alerts into hourly machine-level feature profiles, and (2) weekly feature aggregation and XGBoost model training.

\begin{table}[htbp]
\centering
\caption{Approximate computational requirements per 1,000 machines.}
\label{tab:computational_costs}
\setlength{\tabcolsep}{2.5pt} 
\renewcommand{\arraystretch}{1.05}
\begin{tabular}{@{}lcc@{}}
\toprule
\textbf{Resource} & \textbf{Stream processing} & \textbf{Weekly training} \\
\midrule
CPU (\% of one core)    & 6.3   & 15.0 \\
Peak RAM (MB)           & 83    & 550 \\
Weekly storage (MB)     & 12.85 & 1.68 \\
\bottomrule
\end{tabular}
\end{table}

\section{Discussion}
\label{Sec: Discussion}

\rev{Compared with scenario-based FAIR analyses, LEFSA operationalizes LEF estimation by predicting general machine-level CLE probabilities directly from cybersecurity telemetry and aggregating them across infrastructure layers. It also extends existing organization-level cyber-incident forecasting and machine-level malware prediction by providing risk estimates for general CLEs at all IT infrastructure layers.}

The empirical evaluation demonstrated that machine-level CLE prediction using operational cybersecurity telemetry is feasible, achieving strong discrimination performance and well-calibrated probability estimates across evaluation periods.

The results suggest that machine-level telemetry contains substantial predictive information for future CLE occurrence. In particular, the strong importance of the lagged \rev{outcome} variable indicates temporal dependence in machine-level CLE risk. This suggests that machine-level CLE occurrence may exhibit non-memoryless temporal dynamics, limiting the suitability of simple Poisson-based formulations that assume conditionally independent events with constant intensities. 

The observed overdispersion and residual machine-level dependencies further suggest that infrastructure-level cyber risk aggregation may require explicit modeling of dependence structures rather than assuming conditional independence between machines. Since CLEs are rare and organizations often have many machines, the aggregation of weakly dependent machine-level predictions can amplify dependence effects at the group-level \cite{micro_correlations}. As a result, the independence assumption may underestimate aggregate variance and tail risk.

Among the evaluated models, XGBoost consistently achieved the strongest predictive performance across evaluation periods. The model also demonstrated relatively strong probabilistic calibration, suggesting that machine-level CLE probabilities can be estimated with sufficient reliability to support downstream probabilistic aggregation and cyber risk estimation.

The observed advantage of joint training suggests that machine-level CLE prediction benefits from shared cross-organizational telemetry patterns. This highlights the potential value of MDR-scale datasets and federated learning approaches for operational cyber risk quantification.

Predictive performance and calibration were generally strong and stable across evaluation periods. However, one evaluation period exhibited substantially reduced performance and degraded calibration. This likely reflects a temporary distribution shift in attacker behavior, organizational infrastructure, or operational security processes. This highlights the dynamic nature of cybersecurity environments and emphasizes the importance of continuous retraining and monitoring a LEF model in operational deployments.

Although the empirical evaluation focused on \rev{MDE} telemetry, the LEFSA framework itself is designed to be platform-agnostic and extensible to other telemetry providers and cybersecurity environments. More generally, the framework provides a structured methodology for transforming operational cybersecurity data into probabilistic infrastructure-level cyber risk estimates.

\subsection{Limitations \& operational considerations}

Several limitations should be acknowledged. The empirical evaluation was conducted using proprietary MDR telemetry collected from organizations using \rev{MDE}. Consequently, the dataset cannot be publicly shared, limiting direct reproducibility of the experimental results. The observation window consisted of only 12 weeks, while CLEs remain relatively rare events. Although the machine-level dataset contained a large number of observations, the limited time horizon restricts the ability to evaluate long-term temporal dynamics and rare large-scale incidents.

The empirical evaluation focused primarily on machine-level LEF estimation. While the paper proposed a generalized framework for infrastructure-level aggregation under dependence, accurate modeling of dynamic machine-level dependencies remains a challenging open problem. 

\rev{The empirical evaluation also has several methodological limitations. CLE labels are derived from security incident cases and may not always represent confirmed compromises, while missed detections may introduce systematic label bias. Predictions and aggregate estimates cover only machines observed through the available telemetry and detection platforms, meaning that unmonitored or newly introduced machines are therefore automatically excluded.}

\rev{Operational deployment at scale introduces several additional challenges. LEFSA depends on sufficiently complete and comparable endpoint telemetry, vulnerability information, and detection alerts. Thus machines without adequate monitoring are excluded, potentially causing aggregate risk to be underestimated. Changing machine populations and inconsistent identifiers further complicate tracking and predictions for newly observed assets. A production implementation must also process high-volume data streams while remaining robust to duplicate, delayed, or missing events and service outages. Finally, in the case of cross-organizational model training, it requires appropriate contractual, privacy, governance, and system-integration controls.}

\section{Conclusion \& Future Work}\label{Sec: Conclusion}

This paper introduced LEFSA, a generalized probabilistic framework for estimating LEF through machine-level CLE prediction and hierarchical infrastructure-level aggregation. Rather than relying primarily on manually constructed threat scenarios or externally observable organizational characteristics, LEFSA reformulates LEF estimation as a measurable machine-level prediction problem using operational cybersecurity telemetry.

The empirical evaluation demonstrated that machine-level CLE prediction is feasible using real-world cybersecurity data. XGBoost achieved strong predictive and calibration performance across the evaluation periods. The results further suggest that machine-level telemetry contains substantial predictive information relevant for future CLE occurrence.

Beyond empirical evaluation, the primary contribution of this work is the proposed probabilistic aggregation framework. This framework suggests aggregating calibrated machine-level CLE probabilities across infrastructure layers while preserving uncertainty and dependence structures. This provides a foundation for infrastructure-level cyber risk quantification, including estimation of aggregate CLE distributions, tail risks, and operational cyber resilience metrics.

Several directions for future research remain. Most importantly, future work should investigate dependence-aware aggregation methods capable of modeling dynamic dependencies between machines. In particular, dynamic copula-based approaches in a discrete setting may provide a promising framework for combining marginal machine-level probabilities with time-varying joint dependence structures. Although a single lagged outcome variable was included, future research should investigate richer temporal dependence structures within individual machines, as well as richer telemetry sources, weak supervision, and infrastructure-aware graph representations. Additionally, federated learning could enable training across organizations while preserving data confidentiality and operational privacy constraints. \rev{Potentially, LEFSA could be adjusted to predict future operational incidents.} Finally, integrating LEF estimation with probabilistic loss magnitude models would enable fully quantitative cyber risk estimation aligned with broader FAIR-style risk frameworks.

Overall, this paper suggests that combining probabilistic machine-level CLE models with dependence-aware aggregation is a promising direction for operational and data-driven cyber risk management.

\printbibliography

\end{document}